\documentclass[manuscript]{aastex}

\shorttitle{Noether Gauge Symmetry in Quintom}
\shortauthors{Aslam et al.}

\begin{document}

\title{Noether Gauge Symmetry Approach in Quintom Cosmology}
\author{Adnan Aslam\altaffilmark{1}  Mubasher Jamil\altaffilmark{1} Davood Momeni \altaffilmark{2} \and 
 Ratbay Myrzakulov \altaffilmark{2}  %
Muneer Ahmad Rashid\altaffilmark{1}  Muhammad Raza \altaffilmark{3}}

\begin{abstract}
In literature usual point like symmetries of the Lagrangian have been
introduced to study the symmetries and the structure of the fields. This
kind of Noether symmetry is a subclass of a more general family of
symmetries, called Noether Gauge Symmetries (NGS). Motivated by this
mathematical tool, in this article, we discuss the generalized Noether symmetry
of Quintom model of dark energy, which is a two component fluid model of
quintessence and phantom fields. Our model is a generalization of the
Noether symmetries of a single and multiple components which have been
investigated in detail before. We found the general form of the quintom
potential in which the whole dynamical system has a point like symmetry. We
investigated different possible solutions of the system for diverse family
of gauge function. Specially, we discovered two family of potentials, one
corresponds to a free quintessence (phantom) and the second is in the form
of quadratic interaction between two components. These two families of
potential functions are proposed from the symmetry point of view, but in the
quintom models they are used as phenomenological models without clear
mathematical justification. From integrability point of view, we found two
forms of the scale factor: one is power law and second is de-Sitter. Some
cosmological implications of the solutions have been investigated.
\end{abstract}




\affil{Center for Advanced Mathematics and Physics (CAMP), National
University of Sciences and Technology (NUST), H-12, Islamabad, Pakistan}

\affil{Eurasian International Center for Theoretical Physics, Eurasian
National University, Astana 010008, Kazakhstan} 

\affil{Department of Mathematics, COMSATS Institute of Information
Technology (CIIT), Sahiwal Campus, Pakistan}

\section{Introduction}

Einstein gravity inspired from the equivalence principle, follows the Mach's
principle as the matter creates the geometry or space time concept. There is
no reasonable and clear motive to believe that Einstein gravity must work
beyond the solar system and compatiable to describe the large scale structure of
the whole universe, as well as the gravity in compact objects and solar system.
Just on an adhoc basis and by assuming that the equivalence principle also
works on large scale, the relativistic cosmology of Einstein gravity has
been constructed. Because gravity as a theory for gravitation treats like a
gauge theory, cosmology based on such a gauge theory of gravity is basically
a highly non-linear system of differential equations. It means even there is
no uniqueness principle or theorem for the solutions of the equations. This
is a weak point or very bad freedom in the theory as a mathematical point of
view. Symmetry is a key point for non linear differential equations. Modern
approach is how to find the general point like symmetries of a given
Lagrangian (holonomic or non-holonomic). This appropriate powerful method to
find and investigate the solutions of linear (non linear) dynamical systems
reduces the numbers of the unknown functions by construction of invariants,
the quantities which remain invariance under gauge transformations.
Cosmological models are dynamical systems with attractors and integrable
families. It means, if we start from an initial value of fields then the
cosmological time evolutionary scheme, in the infinite time domain,
asymptotically tends to a state which is free from the initial conditions.
The generic behavior of this type also appears in the other models of
cosmology based on modifications of Einstein gravity.

Noether symmetry is a point like symmetry of Lagrangian and is defined on a
tangent space of the configuration coordinates and their adjoint momenta.
Recently, it has been applied widely in Cosmology from modified curvature
gravities to Einstein-Cartan theories (see \citep{Capozziello:2012hm} for a
brief review).

The motivation to apply Noether symmetry for cosmological models is that in scalar fluid models of dark exotic fluid, quintessence, phantom or quintom, the potential function
is not a unique and known function of fields. In fact, to choose a suitable
and physically acceptable potential function, we have two major approaches:
One is to select a potential function by reconstruction of a special
cosmological behavior following a well known model like LCDM or phantom
fluids. This is reconstruction method and can be used as a useful tool for
fixing the potential function. Second method is that we have used some
phenomenological facts invited from high energy physics. For example, Higgs
bosonic models have been used in inflationary scenarios and the potential
function is in the form that it has spontenaous symmetry breaking and also
renormalizable in the quantum model. Beyond these two major links to the
potential function in scalar field models, the better approach is to find
the potential function by restriction of a general symmetry. Any symmetry in
a Lagrangian defines a conserved charge and from these conservation laws, we
can reduce the numbers of unknown functions, especially, we are able to find
the explicit forms of potential function which posses a definite symmetry.
Noether symmetry or a more generalized form of it, Noether gauge symmetry
gives us such opportunity to reduce the system  to
a first order system of partial differential equations (PDEs) for
generators. The solution of such  first order PDEs give us the form of
generators, conserved charges and especially the form of the interaction
potential function.

Back to the cosmology, we know that numerous astrophysical observations
indicate that the observable universe is undergoing accelerated expansin. This feature has not
a unique full consistent description as a theoretical or phenomenological
model. Another associated startling feature of this expansion is that the
state parameter of dark energy is dynamical which progressively evolves from
sub-negative values to super-negative values (i.e. cross of cosmological
constant boundary $w=-1$), a phenomenon called the Phantom Crossing. This
phantom crossing has been reported in many scenarios of cosmology based on
the dark energy. Recently  modified theories of gravity proposed and one can find numerous
theories of modified gravity in the literature, such as $f(R)$ ($R$ is the
Ricci scalar) \citep{fr,fr2}, $f(R,G)$ where $G$ is Gauss-Bonnet invariant , 
$f(R,T)$ gravity ($T$ is the trace of energy-momentum tensor) %
\citep{frt,frt2,frt3}, $f(R,\mathcal{L}_{m})$ gravity where $\mathcal{L}_{m}$
is the matter Lagrangian \citep{frlm,frlm2} and other ones.

Numerous exact solutions of black holes, wormholes and cosmological models
have been found in these theories. Recently, the Noether symmetries of $f(R)$
and $f(\tau)$ gravity (where $\tau$ is the torsion of space) %
\citep{ftau,ftau2,ftau3,ftau4,ftau5,ftau6,ftau7,ftau8} have been
investigated which help in restricting the astrophysically viable forms of
these functions \citep{ns,ns2,ns3,ns4,ns5,ns6}. Speciefically, by adopting
the Noether symmetry approach, power-law forms ($f(R)\sim R^n$ and $%
f(\tau)\sim \tau^m$, where $m$ and $n$ are finite constants of order unity)
appear (see references in \citep{ns,ns2,ns3,ns4,ns5,ns6}).

One of the these alternative models and the most simple ones is scalar field
models, in which we trust to the time evolution of the scalar field as a
solver of the dark energy problem. In the literature, it has been introduced
the idea of Quintom, in agreement with the current observations %
\citep{feng},\citep{elizalde}. Later, their model was constrained and fitted
with the data of microwave background, supernova and galaxy clustering %
\citep{clu,clu2}. The quintom idea was then extended in other gravitational
setups such as loop quantum cosmology, braneworlds, string theory and
Gauss-Bonnet gravity \citep{lqc,lqc2,lqc3,lqc4,lqc5}. Recently, wormhole
solutions have been studied in Quintom scenario \citep{farook}. For a review
on quintom dark energy model, see \citep{Cai:2009zp}. Quintom model has a
generic unknown potential function of double fields which remains as a
phenomenological function and time evolution of the model is not able to
give any information on the form of it. In this paper, we calculate the
Noether gauge symmetries of the quintom Lagrangian in the physical
background of Friedmann-Robertson-Walker (FRW) space-time. We will propose
two cases of potential with this kind of symmetry.

In this paper, our plan is as follows: In section-II, we write down the
Lagrangian, the idea of Noether gauge symmetries and the corresponding
system of equations. In section-III, we solve the system of differential
equations under some assumptions and constraints on the model parameters and
calculate the Noether symmetries and the first integrals. We discuss
stability in section-IV. Finally we conclude in last section.

\section{Model}

We begin from the simple double field quintom action in four dimensions %
\citep{elizalde},\citep{feng}: 
\begin{equation}
S=\frac{1}{2}\int d^4x \sqrt{-g}\Big[R+\phi_{;\mu}\phi^{;\mu}
-\sigma_{;\mu}\sigma^{;\mu}-2V(\phi,\sigma)\Big] .  \label{s}
\end{equation}
Here $g$ denotes the determinant of the Riemannian metric, $\{\phi,\sigma\}$
are the pair of fields and $V(\phi,\sigma)$ is potential function of fields
which is an unknown function of the model. By observational data the spatial
curvature of the space is negligible. So we adopt the spatially flat FRW
metric as the following 
\[
g_{\mu\nu}=diag(1,-a^2(t)\Sigma_3). 
\]
Here $\Sigma_3$ is metric of three dimensional Euclidean space. We can
rewrite the following field equations 
\begin{eqnarray}
2\frac{\ddot a}{a}+(\frac{\dot a}{a})^2=-p, \\
\ddot\phi+3H\dot{\phi}+\frac{dV}{d\phi}=0, \\
\ddot\sigma+3H\dot\sigma-\frac{dV}{d\sigma}=0.
\end{eqnarray}
Here the effective pressure and dark energy read 
\begin{eqnarray}
\rho=\frac{1}{2}\dot\phi^2-\frac{1}{2}\dot\sigma^2+V(\phi,\sigma), \\
p=\frac{1}{2}\dot\phi^2-\frac{1}{2}\dot\sigma^2-V(\phi,\sigma).
\end{eqnarray}
The point like Lagrangian of (\ref{s}) is 
\begin{equation}
\mathcal{L}(a,\dot a,\phi,\dot \phi,\sigma,\dot\sigma)=-3a\dot a^2+a^3(\frac{%
1}{2}\dot\phi^2-\frac{1}{2}\dot\sigma^2-V(\phi,\sigma)) .  \label{L}
\end{equation}
Lagrangian is holonomic and it posses time translation as trivial symmetry.

We define a vector field 
\begin{equation}
\mathbf{X}=\mathcal{T}\frac{\partial}{\partial t}+\alpha\frac{\partial }{%
\partial a}+\beta\frac{\partial }{\partial \phi}+\gamma\frac{\partial }{%
\partial\sigma}
\end{equation}
is a Noether gauge symmetry (NGS) of the Lagrangian, if 
\begin{equation}
\mathbf{X}\mathcal{L}+\mathcal{L}D_{t}\mathcal{T}=D_{t}G,  \label{NS}
\end{equation}
where the coefficients $\mathcal{T}, \alpha, \beta,\gamma $ and gauge
function $G$ (all functions of $(a,\phi,\sigma)$) are determined from the
Noether symmetry conditions.

Symmetries of the Lagrangian (Noether symmetries), also called the
symmetries of the action integral, are very important as these give double
reduction in the order of the corresponding Euler-Lagrange equation and also
provide conserved quantities \citep{a1,a2,a3,a4}.

Applying (\ref{L}) in (\ref{NS}) we obtain the following system of the PDEs 
\begin{eqnarray}
\mathcal{T}_{a}=0, \mathcal{T}_{\phi}=0, \mathcal{T}_{\sigma}=&&0,  \nonumber
\\
\beta_{\sigma}-\gamma_{\phi}=&&0,  \nonumber \\
6\alpha_{\sigma}-a^{2}\gamma_{a}=&&0,  \nonumber \\
-6\alpha_{\phi}+a^{2}\beta_{a}=&&0,  \nonumber \\
a^{3}\gamma_{t}+G_{\sigma}=&&0,  \nonumber \\
a^{3}\beta_{t}-G_{\phi}=&&0,  \nonumber \\
6a\alpha_{t}+G_{a}=&&0,  \nonumber \\
\alpha+2a\alpha_{a}-a\mathcal{T}_{t}=&&0,  \nonumber \\
3\alpha+2a\beta_{\phi}-a\mathcal{T}_{t}=&&0,  \nonumber \\
3\alpha+2a\gamma_{\sigma}-a\mathcal{T}_{t}=&&0,  \nonumber \\
3a^{2}V\alpha+a^{3}\beta V_{\phi}+a^{3}\gamma V_{\sigma}+a^{3}V\mathcal{T}%
_{t}+G_{t}=&&0 .  \label{sys}
\end{eqnarray}
This is a linear system of partial differential equations. The interaction potential $%
V(\phi, \sigma)$  will be determined by the
above partial differential equations. This potential is in fact a
phenomenological form with parameters which can be adjusted using the data
and in favor of fitting to the astrophysical values. As we know the form of $%
V(\phi, \sigma)$ is just a typical function and in a classical level it has
any arbitrary form. Restriction of the form of $V(\phi, \sigma)$ is one of
the most important results of this paper. We will show how NGS helps us to
fix the form of $V(\phi, \sigma)$ without any reference to the
phenomenological facts.


\section{Solutions}

Finding all solutions of (\ref{sys}) is a hard job. Although, we have
constraints on the forms of the functions but the number of functions and
freedom to fix them is wide. In this section, we will study some particular
solutions of (\ref{sys}) in different cases. In each case, we will find the
corresponding generators and show the closed algebra of the generators.

\textbf{Case 1}: For the gauge function to be arbitrary constant, in which the (%
\ref{sys}) is integrable for set of functions to give the following
particular solutions 
\begin{eqnarray*}
G&&=constant \\
V&&=V(\phi,\sigma)=arbitrary
\end{eqnarray*}
This case gives us no information about the generic form of interaction
potential between quintom components. Moreover, in this case explicitly, the
generators can not be obtained directly from (\ref{sys}). In this special
case with a constant gauge, the only possible symmetry 
\begin{equation}
\mathbf{X}=\frac{\partial}{\partial t}.
\end{equation}
It defines the time translation symmetry corresponds to the conservation of
first integral (energy) of system. This case is the minimal symmetry of
system and here we do not gain much more information than before. This
corresponds to the families of potential which people put as adhoc
phenomenological functions. Due to arbitrary of the gauge such models are
gauge invariant in the language of symmetry.

\textbf{Case 2}: As we can check easily, system given by (\ref{sys}) has
another very important solution for potential even when we set gauge as
constant. The second non trivial family is 
\begin{eqnarray}
G&&=constant,  \nonumber \\
V&&=V(\phi,\sigma)=F(\frac{1}{2}c_{1}(\sigma^{2}-\phi^{2})+c_{2}\sigma-c_{3}%
\phi).  \nonumber  \label{vcase2}
\end{eqnarray}
The following form of interaction reported before as a viable model of
acceleration expansion in the frame of quintom models \cite{Cai:2009zp}. It
corresponds to two family of quadratic potentials. In spite of the previous
case, here the dynamical behavior of whole system is determined by a closed
set of equations of motion. The corresponding Noether symmetries are 
\begin{eqnarray*}
\mathbf{X}_{1}&&=\frac{\partial}{\partial t},  \nonumber \\
\mathbf{X}_{2}&&=(c_{1}\sigma+c_{2})\frac{\partial}{\partial \phi}
+(c_{1}\phi+c_{3})\frac{\partial}{\partial \sigma}.  \nonumber \\
\end{eqnarray*}
These symmetry generators form the simple commutative algebra  
\begin{equation}
\left[\mathbf{X}_{1},\mathbf{X}_{2}\right]=0.
\end{equation}
This family has the crossing phantom line, stable attractors and also a
gauge invariance description. We will study it more later.

\textbf{Case 3}: With constant gauge there also exists another family, in
which we treat one scalar field to be free and another to move in a specific
potential field. The family of solutions here refers to a noninteractive
quintom models, and also, the generic form of the interaction as a single
value function remains undetermined. Just to report the result we write here
\begin{eqnarray}
G&&=constant,  \nonumber \\
V&&=V(\phi,\sigma)=F(\phi).  \nonumber
\end{eqnarray}
One possibility is to take $F(\phi)\sim\phi^2$ and investigate the dynamical
behavior of fields.\newline
In this case, the corresponding Noether symmetries are 
\begin{eqnarray*}
\mathbf{X}_{1}&&=\frac{\partial}{\partial t},  \nonumber \\
\mathbf{X}_{2}&&=\frac{\partial}{\partial \sigma},  \nonumber \\
\end{eqnarray*}
that form the commutative algebra  
\begin{equation}
\left[\mathbf{X}_{1},\mathbf{X}_{2}\right]=0.
\end{equation}
The case if on a free phantom, a constraint quintessence field is subjected
to an unknown potential function $F(\phi)$. Due to the free phantom $\sigma$
it is not very interesting as a double scalar field model.

\textbf{Case 4}: One special non-trivial gauge is that when we consider a
time translational (boost) symmetry through the gauge function. The simple
linear time translation admits the following form of gauge and a non trivial
interaction function  
\begin{eqnarray}
G&&=c_{1}t+c_{2},  \nonumber \\
V&&=V(\phi,\sigma)=-c_{1}\phi+F(\sigma).  \nonumber
\end{eqnarray}
The first term in interaction potential acts as a linear term and has no
non-trivial contribution to the dynamical behavior of the system. Although
it exerts a constant force field but we can absorb it inside the fields
equation by redefining the scalar field as $\phi\rightarrow\phi+c_1t^2/2$.
In this case, the Lagrangian becomes time dependent and the Hamiltonian is
not conserved. It implies a friction in the system due to the existence of a
time dependence Hamiltonian. Here the corresponding Noether symmetry is 
\begin{equation}
\mathbf{X}=\frac{\partial}{\partial t}+\frac{1}{a^{3}}\frac{\partial}{%
\partial \phi}.
\end{equation}

\textbf{Case 5}:  By a weak rotation scheme if we change $t\rightarrow\phi,$
another non zero gauge has been obtained as the following  
\begin{eqnarray}
G&&=c_{1}\phi+c_{2},  \nonumber \\
V&&=V(\phi,\sigma)=F(\sigma).  \nonumber
\end{eqnarray}
It defines a single mode dynamical quintom model and also, here we cannot
fix the interaction function. We have a gauge freedom and one possibility is
to obtain it by taking $F(\sigma)\sim\sigma^2$. The corresponding Noether
symmetries are 
\begin{eqnarray*}
\mathbf{X}_{1}&&=\frac{\partial}{\partial t},  \nonumber \\
\mathbf{X}_{2}&&=(c_{1}\frac{t}{a^{3}}+F(a))\frac{\partial}{\partial \phi}, 
\nonumber \\
\end{eqnarray*}
and the corresponding Lie algebra is  
\begin{equation}
\left[\mathbf{X}_{1},\mathbf{X}_{2}\right] =c_{1}\frac{1}{a^{3}}\frac{%
\partial}{\partial \phi}.
\end{equation}
This case is the $\sigma\rightarrow\phi$ version of the case 4. By the same
reason we are not interesting to it.

\textbf{Case 6}: The case of constant gauge has the following possibility: 
\begin{eqnarray}
G &=&constant,  \nonumber \\
V &=&V(\phi ,\sigma )=constant.  \nonumber
\end{eqnarray}%
It corresponds to two free scalar degrees and cannot explain the
cosmological behavior of the model in accelerated expansion era. We have the
corresponding Noether symmetries: 
\begin{eqnarray*}
\mathbf{X}_{1} &=&\frac{\partial }{\partial t}, \\
\mathbf{X}_{2} &=&\sigma \frac{\partial }{\partial \phi }+\phi \frac{%
\partial }{\partial \sigma }, \\
\mathbf{X}_{3} &=&\frac{\partial }{\partial \sigma }, \\
\mathbf{X}_{4} &=&F(a)\frac{\partial }{\partial \phi }. \\
&&
\end{eqnarray*}%
The closed complete commutative algebra corresponding to these Noether
symmetries is 
\begin{equation}
\left[ \mathbf{X}_{1},\mathbf{X}_{2}\right] =\left[ \mathbf{X}_{1},\mathbf{X}%
_{3}\right] =\left[ \mathbf{X}_{1},\mathbf{X}_{4}\right] =\left[ \mathbf{X}%
_{2},\mathbf{X}_{4}\right] =\left[ \mathbf{X}_{3},\mathbf{X}_{4}\right] =0, 
\nonumber \\
\end{equation}%
\begin{equation}
\left[ \mathbf{X}_{2},\mathbf{X}_{3}\right] =-\frac{\partial }{\partial \phi 
}.
\end{equation}


\section{Exact solutions}


In a dynamical system, when the symmetries are fixed by a tool like NGS,
which we applied here, the next reasonable question is how to find the exact
solutions for different fields using the symmetry generators. Especially
Noether symmetries are useful when we can find the interaction potential
functions of the system. If the generators form a complete non commutative
and fully associative algebra, it is possible to find exact solutions due to
the wide class of symmetries of the original Lagrangian. In our case,
because the system has six different families of symmetries, so we only
examine two simple cases to find exact solutions for scale factor and fields.

For this purpose, we find the invariants and scale parameter for the case
(4) for $c_{1}=0$. As, there is only one Noether symmetry, so the
corresponding conserved quantity or invariant is 
\begin{eqnarray}
\mathbf{I}&=&3a\dot{a}^{2}-\frac{1}{2}a^{3}\dot{\phi}^{2} +\frac{1}{2}a^{3}%
\dot{\sigma}^{2}-a^{3}F(\sigma)+\dot{\phi}.  \label{inter1} \\
\end{eqnarray}
Now the field equations (2)-(4) for the Lagrangian (7) for the Case-4 are 
\begin{eqnarray*}
&&3\dot{a}^{2}+3a^{2}\left(\frac{1}{2}\dot{\phi}^{2} -\frac{1}{2}\dot{\sigma}%
^{2}-F(\sigma)\right)+6a\ddot{a}=0,  \nonumber \\
&&\ddot{\phi}+3\frac{\dot{a}}{a}\dot{\phi}=0,  \nonumber \\
&&\ddot{\sigma}+3\frac{\dot{a}}{a}\dot{\sigma}-\frac{dF(\sigma)}{d\sigma}=0.
\end{eqnarray*}
$\mathbf{I}$ is a conserved quantity, so it must be equal to some constant $c
$ say. Field equations (2)-(4) must satisfy this constraint. We have the
following solutions of the field equations (2)-(4), for the constant $c=0$ 
\begin{eqnarray*}
\phi &=& c_{3} \\
a &=& c_{2}\exp({\pm \frac{\sqrt{6}}{3}t}) \\
F &=& 2 \\
\sigma &=&c_{1}
\end{eqnarray*}
and for the constant $c\neq 0$ 
\begin{eqnarray*}
\phi &=& c_{2} \\
a &=& \frac{1}{2}(6c)^{\frac{1}{3}}t^{\frac{2}{3}} \\
F &=& 0 \\
\sigma &=&\int\frac{\sqrt{2a(c-3a\dot{a}^{2})}}{a^{2}}dt+c_{1},
\end{eqnarray*}
where $c_{1},c_{2}$ and $c_{3}$ are constants. The solutions, which we have
found here, have interesting cosmological implications. For example, the
exponential scale factor denotes a de-Sitter epoch and the power law family
corresponds to a fluid with equation of state of stiff fluid $w=-2$. In the
later case, the explicit form of field $\sigma$ reads as the following: 
\begin{eqnarray}
\sigma(t)=c_1+\frac{2\sqrt{2}}{3}\log t.
\end{eqnarray}
The corresponding Hubble parameter is: 
\begin{eqnarray}
H(t)=\frac{2}{3t}.
\end{eqnarray}
The model mimics LCDM model.


\section{Cosmography}

The form of the quintom interaction given in Case2, is very interesting. In
this section, we want to find the full numerical time evolutionary scheme of
the model, with this potential. Before the full cosmography analysis based
on different scenarios for evolution of the universe in an accelerating
universe has been investigated\citep{bamba}. Attractors and cosmological
predictions have been investigated in details. As a very special case, due
to the specific form of the potential function which we obtained by NGS, we
will study cosmological predictions of our restricted model as a special
two fluids scenario for dark energy. So, in our case we have the explicit
form of $V(\phi ,\sigma )$, so the numerical analysis is done more easily.
In our case, specifically we want to know, how the scale factor,
quintessence and phantom fields evolve in time when we fix the form of the
potential as Case2, by symmetry. Further, as a cosmological result, we want
to study the behaviors of deceleration parameter and EoS parameter, given by 
$w_{eff}=\frac{p_{eff}}{\rho _{eff}}$ as a function of time. In (\ref{vcase2}%
) we make choice $F=X$ as an arbitrary function, also for numerical reasons
we choose $c_{1}=1,c_{2}=c_{3}=\frac{1}{2}$, so that the potential becomes 
\[
V(\sigma ,\phi )=\frac{1}{2}(\sigma -\phi )(\sigma +\phi +1).
\]

\begin{figure*}[thbp]
\begin{tabular}{rl}
\includegraphics[width=7.5cm]{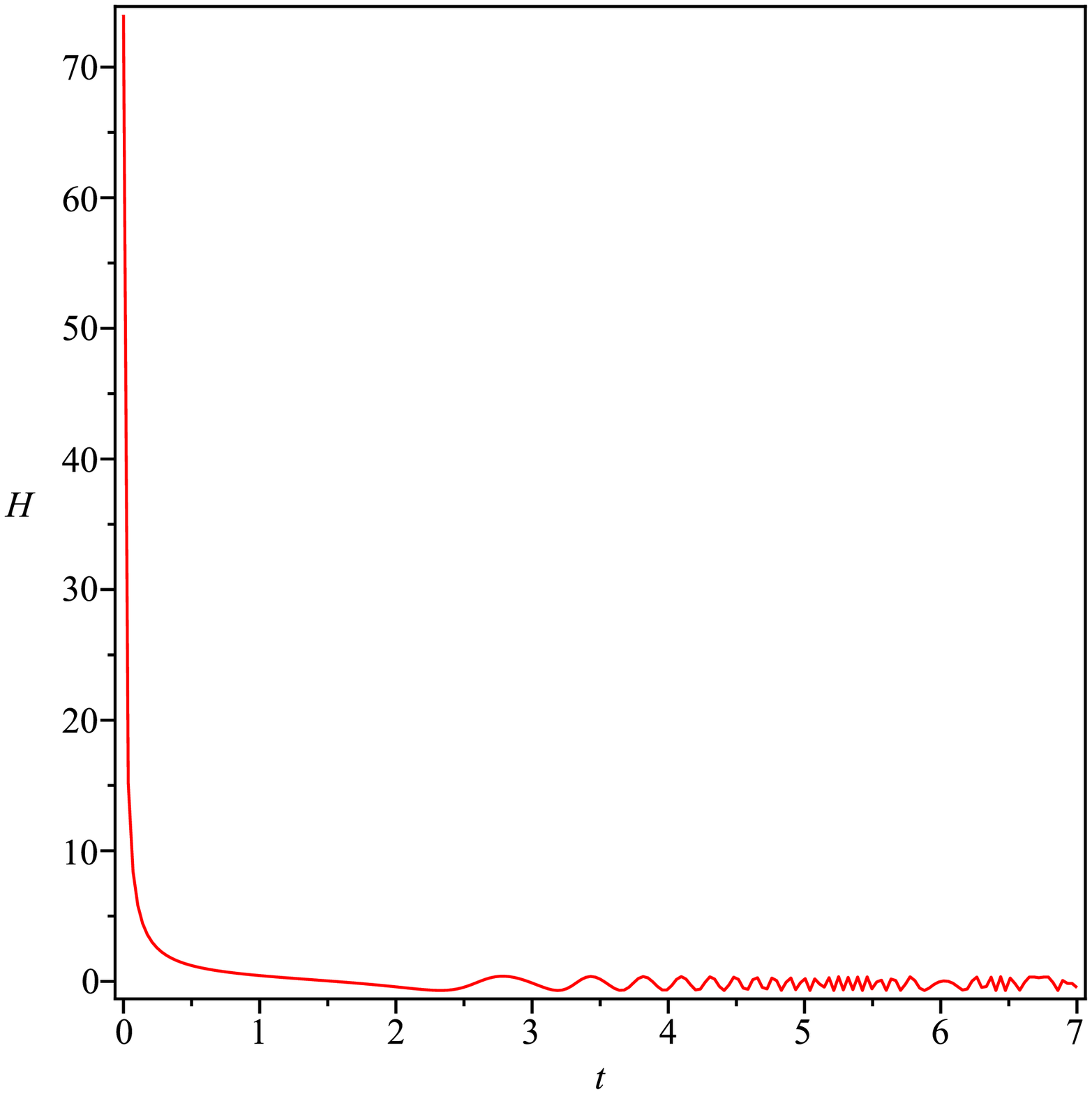} & %
\includegraphics[width=7.5cm]{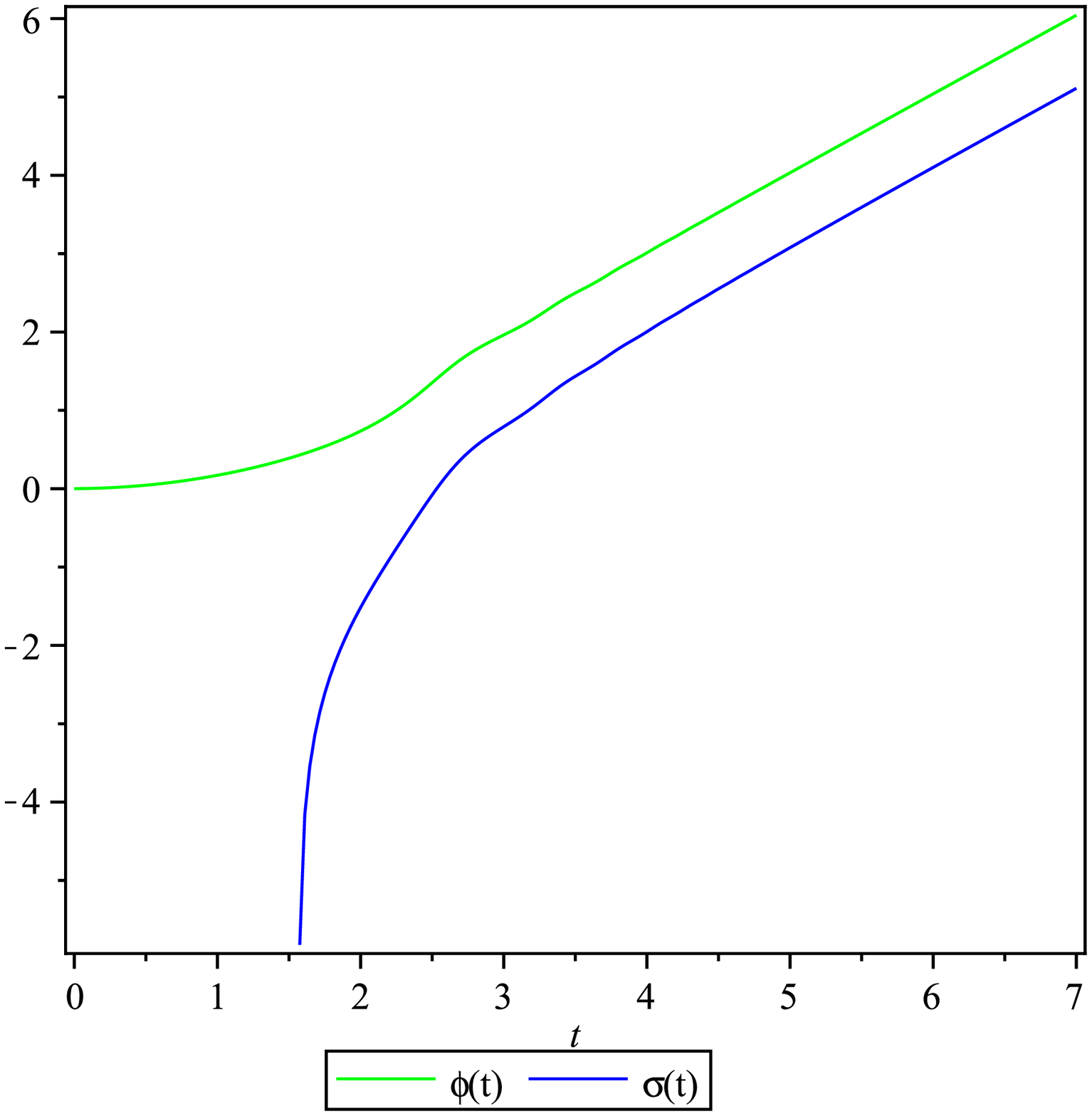} \\ 
\includegraphics[width=7.5cm]{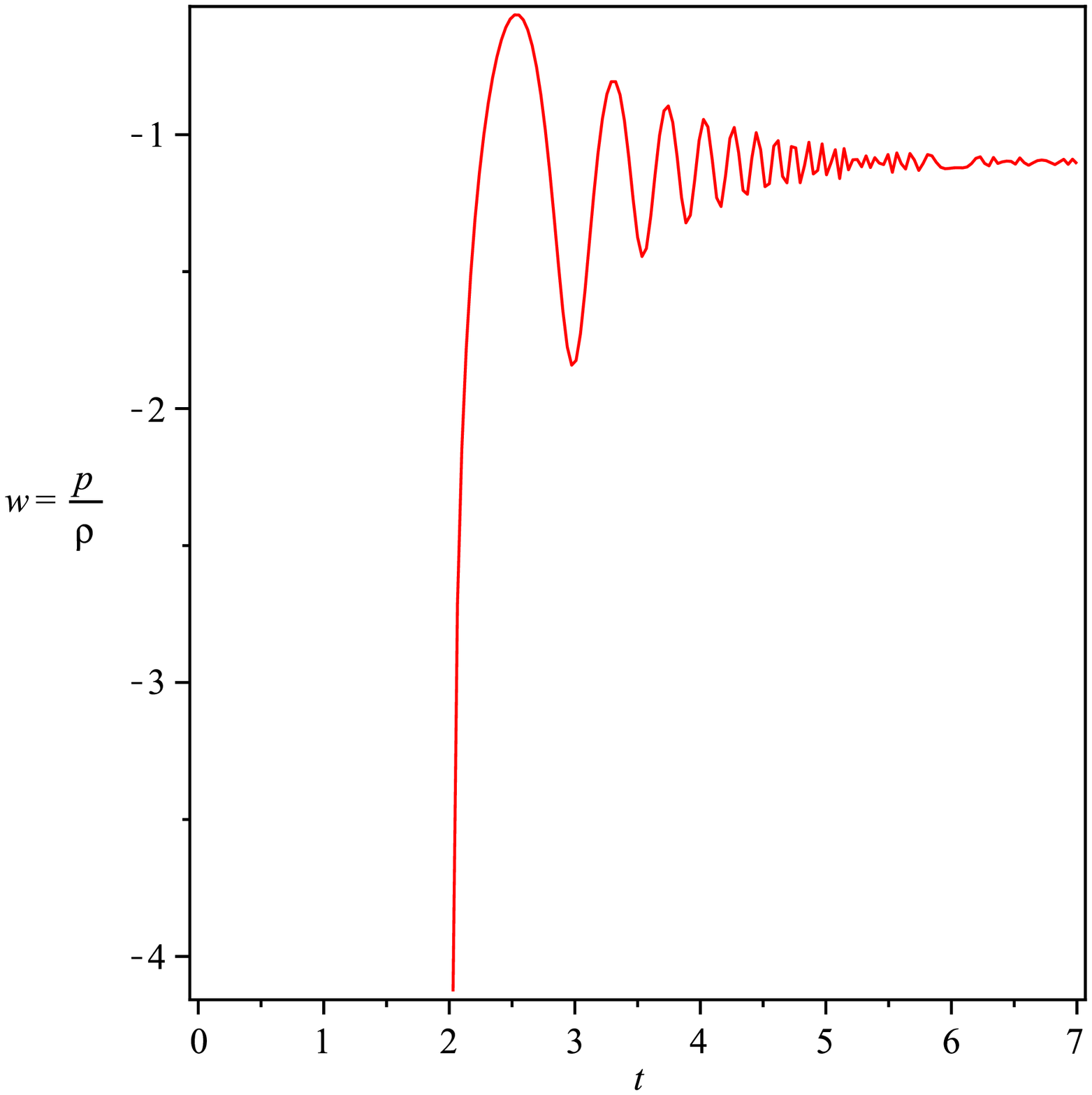} & \includegraphics[width=7.5cm]{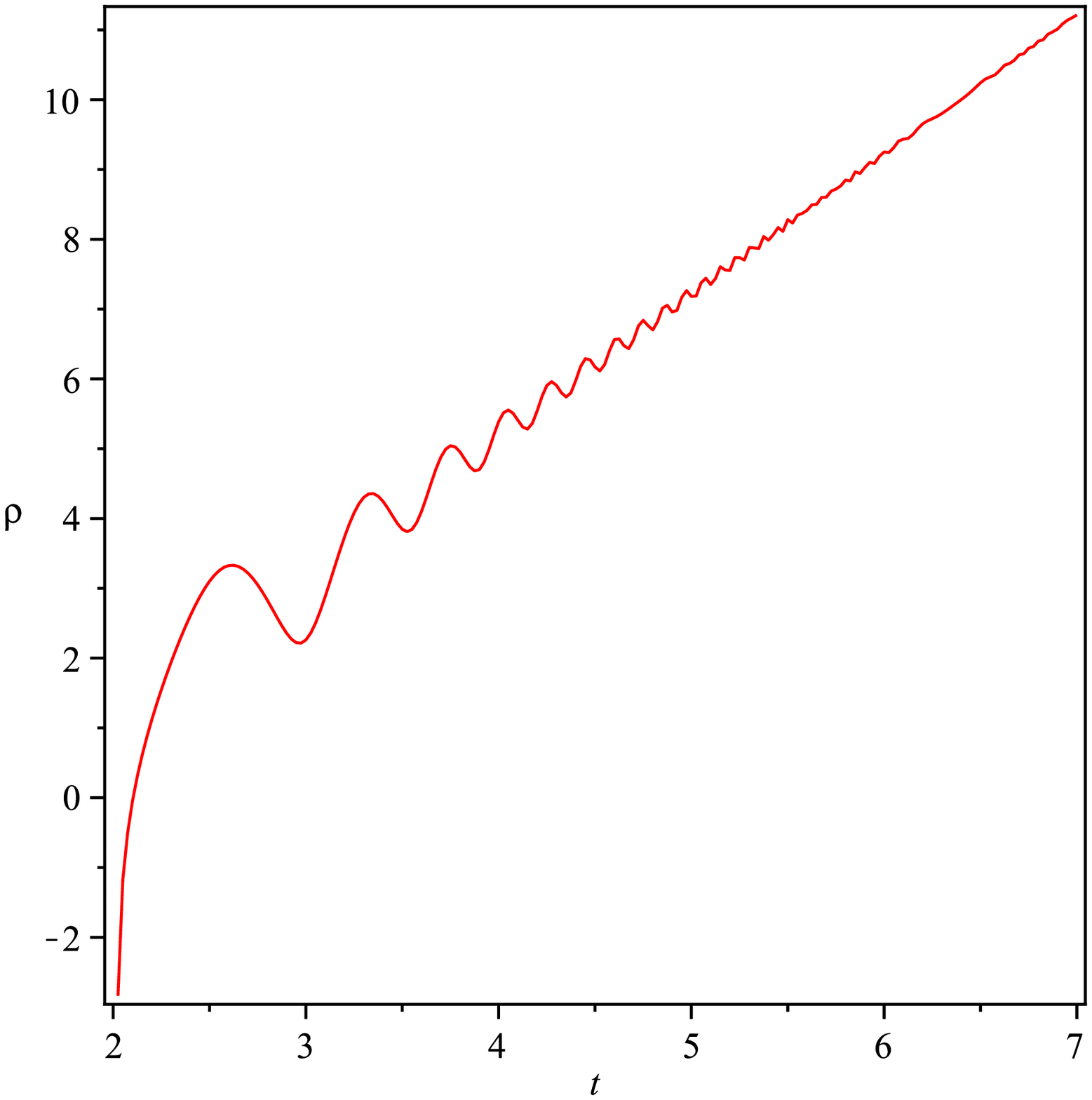}%
\end{tabular}%
\caption{ (\textit{Top Left}) Plot of $H$ vs $t$. It resembles LCDM model. (%
\textit{Top Right})Plot of $\Big[\protect\phi,\protect\sigma\Big]$ vs $t$. (%
\textit{Middle Left}) Plot of effective EoS $w$ vs $t$. It denotes phantom
cross line clearly. (\textit{Middle Right}) Plot of energy density $\protect%
\rho$ vs $t$ for fields. }
\end{figure*}

The system of equations of motion becomes 
\begin{eqnarray}
\ddot{\phi}+3H\dot{\phi}-\frac{1}{2}(1+\phi)=0, \\
\ddot{\sigma}+3H\dot{\sigma}-\frac{1}{2}(1+\sigma)=0, \\
2\dot{H}+3H^2+\frac{1}{2}(\dot{\phi}^2-\dot{\sigma}^2)-\frac{1}{2}%
(\sigma-\phi)(\sigma+\phi+1)=0,
\end{eqnarray}
which can be integrated numerically for a set of functions $\Big[%
H,\phi,\sigma\Big]$. We put the following initial condition: $t = 0, H(t) =
H_0, \phi(0) = 1, \dot{\phi}(0) = 0.2, \sigma(0) = -0.2, \dot{\sigma}(0) = -1
$.

Here we give the illustration of graphs:

From Fig. 1, we deduce that the behavior of Hubble parameter
closely mimics the standard cold dark matter model with vacuum energy. The
Hubble parameter has the highest value in the early Universe while it
suddenly decreases.

In Fig. 2, we plot $\phi$ and $\sigma$ over cosmic time. Asymptotically
their evolution becomes the same, while for small values of time, $\phi$
approaches zero (initial state) while $\sigma$ approaches -5.

We plotted the equation of state parameter against cosmic time in Fig. 3.
Now this $w$ parameter has contributions from the pressure and energy
density of both quintessence and phantom fields. For the suitable choice of
parameters, it appears that $w$ begins to evolve from a highly negative
value and gradually rising to positive values. During this stage, it
obviously crosses the $w=-1$ boundary. It turns out that $w$ follows an
oscillatory behavior which gradually diminishes until $t=7$. We remark that
this value of time does not coincide with the present time since the choice
of parameters is kept arbitrary. It is noticeable that $w$ remains less than 
$-1$ at large time scales.

From Fig. 4, we observe that the total conserved  energy density increases with time
globally. However locally, there are fluctuations in the energy density over
time. It turns out that energy density increases when the phantom scalar
field dominates, while it decreases when the quintessence field decays. For
large cosmic time, the energy density is dominated by the phantom energy. we
also observe that the total energy density increases with time globally.
However locally, there are fluctuations in the energy density over time. It
turns out that energy density increases when the phantom scalar field
dominates while it decreases when the quintessence field decays. For large
cosmic time, the energy density dominates.


\section{Stability}

In double fields model stability problem has been studied before %
\citep{stability}. We are interesting in studying the local stability of the
system. Formally, we want to know that the de-Sitter space is stable or not.
We write the system of the dynamical equations in the following autonomous
system of equations: \par
\begin{eqnarray}
\dot{x}_{1} &=&-3Hx_{1}+\frac{1+\phi }{2}, \\
\dot{x}_{2} &=&-3Hx_{2}+\frac{1+\sigma }{2}, \\
\dot{\sigma} &=&x_{1}, \\
\dot{\phi} &=&x_{2}, \\
\dot{H} &=&-\frac{1}{2}\Big(3H^{2}+\frac{1}{2}(x_{1}^{2}-x_{2}^{2})-\frac{1}{%
2}(\sigma -\phi )(\sigma +\phi +1)\Big).
\end{eqnarray}%
Stationary (critical point) of the system is: 
\[
X_{c}=(x_{1}=0,x_{2}=0,\phi =-1,\sigma =-1,H=0).
\]%
The model with $H=0$ is a sub case of de-Sitter so called as static model or
Einstein universe. We perturb the system around this critical point up to
first order to find: 
\begin{eqnarray}
\delta \ddot{\chi}-\frac{1}{2}\delta \chi  &=&0,\ \ \chi =\sigma -\phi , \\
\delta \dot{H}+\frac{1}{4}\delta \chi  &=&0.
\end{eqnarray}%
General solution of the model is written as the following: 
\[
\delta \chi =\delta \chi _{0}\sin (t/\sqrt{2}+\theta _{0}),\ \ \delta H=%
\frac{\sqrt{2}}{4}\delta \chi _{0}\cos (t/\sqrt{2}+\theta _{0}).
\]%
Oscillatory solutions indicate that the model is stable under small
perturbation in first order. So our quintom scenario predicts the stability
scenario of de-Sitter case.


\section{Conclusions}

To explain the acceleration expansion era of the current universe quintom
model proposed as a double field model, contained of two fields,
quintessence and phantom field. The typical form of the potential function $%
V(\sigma ,\phi )$ remains as a phenomenological unknown function. In this
article, we studied Noether gauge symmetry of quintom cosmology with
arbitrary potentials via Noether Gauge symmetry, as a generalization of the
popular Noether symmetry. Such an approach has been previously investigated
in the quintom cosmology for dynamical system analysis. We presented the
Noether symmetry generators for a typical interactive quintom model. In
general, we obtained six families of quintom models based on different
generators. In case 1, we took the gauge field constant and with the time
translation symmetry. Family 2, corresponded to two families of quadratic
potentials where the dynamical behavior of whole system determined by a
closed set of equations of motion. In this family $V(\phi ,\sigma )=F(\frac{1%
}{2}c_{1}(\sigma ^{2}-\phi ^{2})+c_{2}\sigma -c_{3}\phi )$. For this case,
we performed cosmography analysis in details. It is interesting to note that
the behavior of Hubble parameter closely mimics the standard cold dark
matter model with vacuum energy. Furthermore, the total energy density
increases with time monotonically, globally. Also, numerically, we deduced
that EoS parameter $w$ began to evolve from a highly negative value and
gradually rising to positive values. Also we showed that Einstein space is
stable as a special case. During this stage, it obviously crosses the $w=-1$
boundary. So, we extend the cosmology of quintom models beyond the
phenomenological potential functions by a new approach of Noether Gauge
Symmetry. Further, as we know the equivalent presentation of quintom theory
gives specific fluid with a pair of effective energy density and pressure.
Consequently, the exact solutions for the generators and the potential
function found, remain to be solutions for field equations in the other
equivalent description via fluids with effective quantities. In this new
representation, conservation of energy in terms of the Liouvile's form is $%
\frac{D\rho _{eff}}{Dt}=0$ where it is equivalent to the usual conservation equation $\dot{\rho_{eff}}+3H(\rho_{eff}+p_{eff})=0$. In this equivalent form, NGS of point like scalar field Lagrangian now
represent a kind of the symmetry in fluid description. Existence of a closed
algebra between the generators defines a sub class of Lie symmetries. This
Lie symmetry now transfers to the fluid's equivalent description and so the
corresponding Noether charges (fluid's invariants) remain conservative for
effective fluid. So, the quintom model inspired from the Noether gauge
symmetry is able to give a set of reasonable predictions. Finally, we
mention here that this method may be applied to generalized quintom model
introduced in \citep{gquintom}. In such generalized quintom models by using NGS, we will be able to find explicitly the form of the
potential function . \acknowledgments The work of M. Jamil is supported from the
financial grant No. 20-2166/NRPU/RD/HEC/12-5699 of the Higher Education
Commission, Pakistan.

\end{document}